\documentclass[12pt]{article}
\usepackage{amsmath,graphicx}
\numberwithin{equation}{section}
\usepackage{feynmf}

 \newcommand{\be}{\begin{equation}}
\newcommand{\ee}{\end{equation}}

\newcommand{\beqa}{\begin{eqnarray}}
\newcommand{\eeqa}{\end{eqnarray}}
\newcommand{\nn}{\nonumber}

\newcommand{\<}{\langle}

\def\CA {{\cal A}}

\def\CG {{\cal G}}
\def\CH {{\cal H}}

\def\CL {{\cal L}}

\def\CP {{\cal P}}

\def\CV {{\cal V}}

\begin{document}
\setlength{\unitlength}{1mm}
\setlength{\baselineskip}{7mm}
\begin{titlepage}

\begin{flushright}

{\tt NRCPS-HE-52-2015} \\

\end{flushright}

\vspace{1,5cm}
\begin{center}

{\Large \it  Generalisation of the Yang-Mills Theory
\\
\vspace{0,3cm}

} %title ends

\vspace{1cm}
%\author{
{\sl  George Savvidy

\bigskip
\centerline{Institute of Nuclear and Particle Physics}
\centerline{${}^+$ \sl Demokritos National Research Center, Ag. Paraskevi,  Athens, Greece}
\bigskip
}

\end{center}

 \begin{abstract}
We suggest an extension of the gauge principle which includes tensor gauge fields.
In this extension of the Yang-Mills theory the vector gauge boson becomes a
member of a bigger family of gauge bosons of arbitrary large integer spins.
The proposed extension
is essentially based on the extension of the Poincar\'e 
algebra and the existence of an appropriate transversal representations.
The invariant Lagrangian is expressed in terms of new
higher-rank field strength tensors. It does not contain higher derivatives of
tensor gauge fields and all interactions take place through three- and four-particle
exchanges with a dimensionless coupling constant.
We calculated the scattering amplitudes of non-Abelian tensor gauge bosons
at tree level, as well as their one-loop contribution into the Callan-Symanzik beta function.
This contribution is negative and
corresponds to the asymptotically free theory. Considering the contribution
of tensorgluons of all spins into the beta function we found that it is 
leading to the theory which is  conformally invariant at very high energies.
The proposed extension may
lead to a natural inclusion of the standard theory of fundamental forces
into a larger theory in which vector gauge bosons, leptons and quarks represent
a low-spin subgroup. We consider a possibility
that inside the proton and, more generally, inside hadrons there are
additional partons - tensorgluons, which can carry a part of the proton
momentum. The extension of QCD influences the unification scale at which
the coupling constants of the Standard Model merge, shifting its value to
lower energies.

\end{abstract}

\vspace{1cm}

\centerline{\it Talk given at the "Conference on 60 Years of Yang-Mills Gauge Field Theories"}
\centerline{\it   Singapore 2015}

\end{titlepage}

\newpage

\pagestyle{plain}

\section{\it Introduction}

It is well understood that the concept of local gauge
invariance formulated by Yang and Mills \cite{yang}
allows to define the non-Abelian gauge fields ,
to derive their  dynamical field equations
and to develop a universal point of view on matter interactions
as resulting from the exchange of gauge quanta of different forms.
The fundamental forces - electromagnetic, weak and strong interactions
are successfully described by the non-Abelian Yang-Mills fields.
The vector-like gauge particles - the photon, $W^{\pm},Z$ and
gluons mediate interaction
between smallest constituents of matter - leptons and quarks.

The non-Abelian local gauge invariance, which was formulated
by Yang and Mills \cite{yang},
requires that all interactions must be invariant under
independent rotations of internal
charges at all space-time
points. The gauge principle allows very little arbitrariness: the interaction of matter
fields,  which carry non-commuting internal charges, and the nonlinear
self-interaction of gauge bosons are essentially fixed by the requirement
of local gauge invariance, very similarly to the self-interaction of
gravitons in general relativity \cite{chern,Weyl:1929fm,weyl,Cartan:1923zea,cartan,utiyama,kibble}.

It is therefore appealing to extend the gauge principle, which was elevated by Yang and
Mills to a powerful constructive principle, so that it will define the
interaction of matter fields which carry
not only non-commutative internal charges, but
also arbitrary large spins\footnote{The research in high spin field theories 
has long and rich history.  One should mention the early works of 
Majorana \cite{majorana}, Dirac \cite{dirac}, Fierz \cite{fierz}, Pauli \cite{fierzpauli}, Schwinger \cite{schwinger}, Singh and Hagen \cite{singh}, Fronsdal \cite{fronsdal}, 
Weinberg \cite{Weinberg:1964cn}, Minkowski \cite{minkowski}, Brink  et.al. \cite{Bengtsson:1983pd}, 
Berends, Burgers and Van Dam \cite{berends},
Ginsburg  and Tamm \cite{ginzburg,Ginzburg:1974qz}, Nambu \cite{Nambu:1968sa},
Ramond \cite{Ramond:1971gb}, Brink \cite{Brink:2004db}, Fradkin\cite{fradkin}, 
Vasiliev \cite{vasiliev}, 
Sagnotti, Sezgin and Sundel \cite{Sagnotti:2005ns}, Metsaev \cite{Metsaev:2005ar}, 
Gabrielli \cite{Gabrielli:1990ay}, Castro \cite{Castro:2006kd}, Manvelyan et.al. \cite{Manvelyan:2010jr},   and many other works (see also the references in \cite{Ginzburg:1974qz,vasiliev,Guttenberg:2008qe}). }.  It seems that this will naturally
lead to a theory in which fundamental forces will be mediated by
integer-spin gauge quanta  and
that the Yang-Mills vector gauge boson will become a member of a bigger
family of tensor gauge bosons \cite{Savvidy:2005fi,Savvidy:2005zm,Savvidy:2005ki,
Savvidy:2010vb,Savvidy:2010kw}

The proposed extension of Yang-Mills theory 
is essentially based on the extension of the Poincar\'e 
algebra and the existence of an appropriate transversal representations
of that algebra.  The tensor gauge fields take value in extended 
Poincar\'e algebra. 
The invariant Lagrangian is expressed in terms of new
higher-rank field strength tensors. The Lagrangian  does not contain higher derivatives of
tensor gauge fields and all interactions take place through three- and four-particle
exchanges with a dimensionless coupling constant  \cite{Savvidy:2005fi,Savvidy:2005zm,Savvidy:2005ki,Savvidy:2010vb,Savvidy:2010kw}.

It is important to calculate the scattering amplitudes of non-Abelian tensor gauge bosons
at tree level, as well as their one-loop contribution into the Callan-Symanzik beta function.
This contribution is negative and
corresponds to the asymptotically free theory. The proposed extension may
lead to a natural inclusion of the standard theory of fundamental forces
into a larger theory in which vector gauge bosons, leptons and quarks represent
a low-spin subgroup \cite{Savvidy:2014hha,Savvidy:2013zwa,Savvidy:2014uva}. 

In the line with the above development we considered a possible extension of QCD. 
In so extended QCD the spectrum of the theory contains new bosons, {\it the tensorgluons},
in addition to the quarks and gluons. The tensorgluons have
zero electric charge, like gluons, but have a larger spin. 
Radiation of tensorgluons by gluons leads to a
possible existence of tensorgluons inside the proton and, 
more generally, inside the hadrons. Due to the emission
of tensorgluons part of the proton momentum which is 
carried  by the neutral constituents can be shared between 
gluons and tensorgluons. The density of neutral partons  is
therefore given by the sum: $G(x,t)+T(x,t)$, where $T(x,t)$
is the density of the tensorgluons\cite{Savvidy:2014hha,Savvidy:2013zwa,Savvidy:2014uva}.
 To disentangle these contributions and to 
decide which piece of the neutral partons is the contribution of gluons $G(x,t)$
and which one is of the tensorgluons one should measure the helicities of the neutral
components, which seems to be a difficult task.

The extension of QCD influences the unification scale at which
the coupling constants of the Standard Model merge.
We observed that the unification scale at which
standard coupling constants are
merging is  shifted to lower energies telling us that it may be that a
new physics is round the corner.
Whether all these phenomena are consistent with experiment is an open question.

The paper is organised as follows. In Section 2 we shall define the 
composite gauge field ${\cal A}_{\mu}(x,e)$,  
which depends on the space-time coordinates $x_{\mu}$ and the new space-like vector variable $e^{\lambda}$.   The high-rank tensor gauge fields $A^{a}_{\mu\lambda_1 ... \lambda_{s}}(x)$ appear in the expansion of ${\cal A}_{\mu}(x,e)$ over the vector variable.  
We introduce a corresponding 
extension of the Poincar\'e algebra $L_{G}(\CP)$ and consider the high-rank fields as 
tensor gauge fields taking  value in algebra $L_{G}(\CP)$.   
In Section 3 we shall describe the transversal representation 
$L^{\bot}$ of the  generators of the algebra $L_{G}(\CP)$, their helicity content and their invariant scalar products. The fact that the representation of the generators is transversal plays an important role in the definition of the gauge field $\CA_{\mu}(x,e)$. 
In transversal representation the tensor gauge fields are 
projecting out into the plane transversal to the momentum and contain only 
positive  space-like components of a definite helicity. 

In Section 4 we shall define the gauge 
transformation of the gauge fields, the field 
strength tensors and the invariant Lagrangian.  The kinetic term describes the propagation
of positive definite helicity states. The helicity spectrum of the propagating modes 
is consistent  with the helicity spectrum which appears in the projection of the tensor 
gauge fields into transversal generators $L^{\bot}$.
The Lagrangian  defines not only a free propagation of tensor gauge bosons, but also their
interactions. The interaction diagrams for the lower-rank bosons are
presented on Fig.\ref{fig1}-\ref{fig2}. The high-rank bosons  interact
through the triple and quartic interaction vertices with a dimensionless coupling constant. 
In Section 5 we shall  calculate and study
the scattering amplitudes of the vector and tensor gauge bosons and their splitting
 amplitudes by using spinor representation of the momenta and polarisation tensors.   

In Section 6 we shall  consider a possibility that inside the proton
and, more generally, inside hadrons there are additional partons - tensorgluons, 
which can carry a part of the proton momentum.  We generalise the DGLAP equation 
which includes the splitting probabilities of the gluons into tensorgluons and calculated 
the one-loop Callan-Simanzik beta function. This contribution is negative and
corresponds to the asymptotically free theory. Considering the contribution
of tensorgluons of all spins into the beta function we found that it is
leading to the theory which is  {\it conformally invariant} at very high energies. 
In Section 7 we observed that the 
unification scale at which standard coupling constants are
merging is  shifted to lower energies.  
In conclusion we summarise the results and discuss 
the challenges of the experimental verification of the suggested model.

\section{\it Tensor Gauge Fields and Extended Poincar\'e Algebra}

The gauge fields are defined as rank-$(s+1)$ tensors
\cite{Savvidy:2005fi,Savvidy:2005zm,Savvidy:2005ki}
\be\label{tensorfields}
A^{a}_{\mu\lambda_1 ... \lambda_{s}}(x),~~~~~s=0,1,2,...
\ee
and are totally symmetric with respect to the
indices $  \lambda_1 ... \lambda_{s}  $.  A priory the tensor fields
have no symmetries with
respect to the first index  $\mu$. The index $a$ numerates the generators $L_a$
of the Lie algebra $L_G$ of a  compact Lie group G with totally antisymmetric
structure constants $f_{abc}$.

The tensor fields (\ref{tensorfields}) can be considered as the components of a composite
 gauge field ${\cal A}_{\mu}(x,e)$ which depends on additional translationally
 invariant space-like unite vector \cite{Savvidy:2005ki,Savvidy:dv,Savvidy:2003fx,Savvidy:2005fe}:
\be\label{unite}
e_{\lambda} e^{\lambda}=-1.
\ee
A similar vector variable, in addition to the space-time coordinate $x$,  was introduced
earlier by Yakawa \cite{yukawa1}, Fierz \cite{fierz1}, Wigner\cite{wigner}, Ginzburg and Tamm \cite{ginzburg,Ginzburg:1974qz} and
others  \cite{vasiliev}.   The  variable $e^\lambda$ is  also
reminiscent  to the Grassmann variable $\theta$ in supersymmetric
theories where the superfield $\Psi(x,\theta)$ depends on two variables $x$ and $\theta$
\cite{Golfand:1971iw,Salam:1974pp}.
We shall consider all tensor gauge fields (\ref{tensorfields}) as the components
appearing  in the expansion over the above mentioned
vector variable \cite{Savvidy:2005ki}:
\be\label{gaugefield}
{\cal A}_{\mu}(x,e)=\sum_{s=0}^{\infty} {1\over s!} ~A^{a}_{\mu\lambda_{1}...
\lambda_{s}}(x)~L_{a}e^{\lambda_{1}}...e^{\lambda_{s}}.
\ee
The gauge field $A^{a}_{\mu\lambda_1 ... \lambda_{s}}$ carries
indices $a,\lambda_1, ..., \lambda_{s}$ which are labelling the
generators  $L_{a}^{\lambda_1 ... \lambda_{s}} = L_a e^{\lambda_1}...e^{\lambda_s}$
of extended current
algebra $L_{\CG}$ associated with the  Lie algebra $L_G$ \cite{Savvidy:2005ki,Savvidy:2010vb}.
The algebra $L_{\CG}$
 has infinitely many generators
$L_{a}^{\lambda_1 ... \lambda_{s}} $ and
 is given by the commutator \cite{Savvidy:2005ki,Savvidy:2010vb,Savvidy:2010kw}
\be\label{currentalge}
[L_{a}^{\lambda_1 ... \lambda_{k}}, L_{b}^{\lambda_{k+1} ... \lambda_{s}}]=if_{abc}
L_{c}^{\lambda_1 ... \lambda_{s}}~,~~~~~s=0,1,2....
\ee
The generators $L_{a}^{\lambda_1 ... \lambda_{s}}$ commute to themselves forming an infinite series of commutators of current algebra $L_{\CG}$ which
cannot be truncated, so that the index s runs from zero to infinity.
Because the generators $L_{a}^{\lambda_1 ... \lambda_{s}}$ are
space-time tensors, the full algebra should include the Poincar\'e generators $P^{\mu},~M^{\mu\nu}$ as well. This naturally leads to the
extension $L_G (\CP )$ of the Poincar\'e algebra $L_{\CP}$
\cite{Savvidy:2010vb,Savvidy:2010kw,Antoniadis:2011re}:
\beqa\label{extensionofpoincarealgebra}
~&&[P^{\mu},~P^{\nu}]=0,\nn\\
~&&[M^{\mu\nu},~P^{\lambda}] = \eta^{\nu\lambda }~P^{\mu}
- \eta^{\mu\lambda  }~P^{\nu} ,\nn\\
~&&[M^{\mu \nu}, ~ M^{\lambda \rho}] = \eta^{\mu \rho}~M^{\nu \lambda}
-\eta^{\mu \lambda}~M^{\nu \rho} +
\eta^{\nu \lambda}~M^{\mu \rho}  -
\eta^{\nu \rho}~M^{\mu \lambda} ,\nonumber\\
~&&[P^{\mu},~L_{a}^{\lambda_1 ... \lambda_{s}}]=0, \nn\\
~&&[M^{\mu \nu}, ~ L_{a}^{\lambda_1 ... \lambda_{s}}] =
\eta^{\nu \lambda_1}~L_{a}^{\mu \lambda_2... \lambda_{s}}
-\eta^{\mu\lambda_1}~L_{a}^{\nu\lambda_2... \lambda_{s}}
+...+
\eta^{\nu \lambda_s}~ L_{a}^{\mu \lambda_1... \lambda_{s-1}} -
\eta^{\mu \lambda_s}~L_{a}^{\nu \lambda_1... \lambda_{s-1}} ,\nonumber\\
~&&[L_{a}^{\lambda_1 ... \lambda_{k}}, L_{b}^{\lambda_{k+1} ... \lambda_{s}}]=if_{abc}
L_{c}^{\lambda_1 ... \lambda_{s}}.
\eeqa
We have here an extension of the Poincar\'e algebra by generators
$L_{a}^{\lambda_1 ... \lambda_{s}}$ which carry the
{\it  internal charges  and spins}.
The algebra $L_G (\CP )$ incorporates the Poincar\'e  algebra $L_{\CP}$ and an
internal algebra $L_G$ in a nontrivial way, which is different from the direct
product.

There is no conflict with the Coleman-Mandula theorem
\cite{Coleman:1967ad,Haag:1974qh} because
the theorem applies to the symmetries that act on S-matrix elements and not on all
the other symmetries that occur in quantum field theory.
The above  symmetry  group (\ref{extensionofpoincarealgebra}) is the
symmetry which acts on the gauge field
$\CA_{\mu}(x,e)$ and is not the symmetry of the S-matrix.
 The theorem
assumes among other things that the vacuum is nondegenerate and
that there are no massless particles in the spectrum. As we shall see, the
spectrum of the extended Yang-Mills theory is massless.

{\it In order to define the gauge field $\CA_{\mu}(x,e)$ in (\ref{gaugefield}) and find out
its helicity content one should specify the representation of the generators
$L_{a}^{\lambda_1 ... \lambda_{s}}$ in algebra (\ref{extensionofpoincarealgebra}).
In the next section we shall describe the
so called transversal representation, which is used to define the tensor gauge fields in  the
decomposition (\ref{gaugefield}).}

\section{\it Transversal Representation of Algebra $L_G (\CP )$}

The important property of the algebra (\ref{extensionofpoincarealgebra}) is its
invariance  with respect to the following "gauge"
transformations \cite{Savvidy:2010vb,Savvidy:2010kw,Antoniadis:2011re}:
\beqa\label{isomorfism}
& L_{a}^{\lambda_1 ... \lambda_{s}} \rightarrow L_{a}^{\lambda_1 ... \lambda_{s}}
+ \sum_{1} P^{\lambda_1}L_{a}^{\lambda_2 ... \lambda_{s}}+
\sum_{2} P^{\lambda_1} P^{\lambda_2} L_{a}^{\lambda_3 ... \lambda_{s}} +...+
P^{\lambda_1}... P^{\lambda_s} L_{a} \nn\\
& M^{\mu\nu} \rightarrow M^{\mu\nu},~~~~
P^{\lambda} \rightarrow P^{\lambda},
\eeqa
where the sums  $\sum_{1},\sum_{2},... $ are over all inequivalent index permutations.
The above transformations contain polynomials of the momentum  operator $P^{\lambda}$
and are reminiscent of the gauge field  transformations.
This is  ``off-shell"  symmetry
because the invariant operator $P^2$ can have any value.
As a result, to any given  representation of
$L_{a}^{  \lambda_1 ... \lambda_{s}},~s=1,2,...$
one can add the longitudinal terms, as it follows from the  transformation (\ref{isomorfism}).
All representations are therefore defined modulo longitudinal terms,
and we can identify these generators as "gauge generators".

The second general property of the extended algebra is that each gauge generator
$ L_{a}^{  \lambda_1 ... \lambda_{s}}$ cannot be
realised as an irreducible representation of the Poincar\'e  algebra of a definite helicity,
i.e. to be a {\it  symmetric and traceless tensor}. The reason is that the commutator of
two symmetric traceless generators in (\ref{extensionofpoincarealgebra}) is not any more
a traceless tensor. Therefore the  generators $L_{a}^{  \lambda_1 ... \lambda_{s}}$
realise a reducible representation of the Poincar\'e algebra and each of
them carries a spectrum of helicities, which we shall describe below.

The algebra $L_G(\CP)$ has representation in terms of differential operators of the
following general form:
\beqa\label{represofextenpoincarealgebra}
~&& P^{\mu} = k^{\mu} ,\nn\\
~&& M^{\mu\nu} = i(k^{\mu}~ {\partial\over \partial k_{\nu}}
- k^{\nu }~ {\partial \over \partial k_{\mu}}) + i(e^{\mu}~ {\partial\over \partial e_{\nu}}
- e^{\nu }~ {\partial \over \partial e_{\mu}}),\nn\\
~&& L_{a}^{\lambda_1 ... \lambda_{s}} =e^{\lambda_1}...e^{\lambda_s} \otimes L_a,
\eeqa
where $e^{\lambda} $ is a translationally invariant space-like unite vector (\ref{unite}).
The vector space of a representation  is parameterised
by the momentum $k^{\mu}$ and translationally invariant
vector variables $e^{\lambda}$:
\be\label{vectorspace}
\Psi(k^{\mu}, e^{\lambda} )~.
\ee
The irreducible representations can be obtained from (\ref{represofextenpoincarealgebra})
by   imposing  invariant constraints on the
vector space of functions  (\ref{vectorspace})
of the following form \cite{wigner,wigner1,yukawa1,fierz1}:
\be\label{constraint}
k^2=0,~~~k^{\mu} e_{\mu}=0,~~~e^2=-1~.
\ee
These equations have a unique solution \cite{wigner}
\be\label{solution}
e^{\mu}= \chi k^{\mu} + e^{\mu}_{1}\cos\varphi +e^{\mu}_{2}\sin\varphi,
\ee
where $e^{\mu}_{1}=(0,1,0,0),~ e^{\mu}_{2}=(0,0,1,0)$ when $k^{\mu}=\omega(1,0,0,1)$.
The  $\chi$ and  $\varphi$ remain as independent variables
on the cylinder $ \varphi \in S^1, \chi \in R^1 $. The invariant subspace of functions (\ref{vectorspace}) now reduces to
the following form:
\be\label{independentvariables}
\Psi(k^{\mu}, e^{\nu} )~\delta(k^2)~\delta(k\cdot e)~\delta(e^2 +1)
= \Phi(k^{\mu}, \varphi, \chi) .
\ee
If we take into account (\ref{solution}) the generators
$L_{a}^{ \lambda_1 ... \lambda_{s}}= e^{\lambda_1}...e^{\lambda_s} \otimes L^a $,
it takes the following form:
\be\label{trasversalgenera}
L_{a}^{\bot~ \lambda_1 ... \lambda_{s}}= \prod^{s}_{n=1} ( \chi k^{\lambda_n} + e^{\lambda_n}_{1}\cos\varphi
+e^{\lambda_n}_{2}\sin\varphi) \otimes L_a.
\ee
This is a purely transversal representation because of (\ref{constraint}):
\be\label{transversality}
k_{\lambda_1}L_{a}^{\bot \lambda_1 ... \lambda_{s}}=0,~~~~s=1,2,...
\ee
The generators
$L_{a}^{\bot~ \lambda_1 ... \lambda_{s}}$  carry helicities in the following range:
\be\label{trasversalgenera1}
h=(s,s-2,......, -s+2, -s),
\ee
in total $s+1$ states.  Indeed,
this can be deduced from the explicit representation (\ref{trasversalgenera}) by using
helicity polarisation vectors $e^{\lambda}_{\pm}= (e^{\lambda}_1 \mp i e^{\lambda}_2)/2$:
\be\label{trasversalgenera11}
L_{a}^{\bot~ \lambda_1 ... \lambda_{s}}= \prod^{s}_{n=1} ( \chi k^{\lambda_n} + e^{i \varphi} e^{\lambda_n}_{+}
+e^{-i \varphi} e^{\lambda_n}_{-})\oplus L_a.
\ee
Performing the multiplication
in (\ref{trasversalgenera11}) and collecting the terms of a given power
of momentum we shall get the following expression:
\beqa\label{trasversalgenera2}
L_{a}^{\bot~ \mu_1 ... \mu_{s}}= \prod^{s}_{n=1} (e^{i \varphi} e^{\mu_n}_{+}
+e^{-i \varphi} e^{\mu_n}_{-})\oplus L_a +~~~~~~~~~~~~~~~~~~~~~~~~~~~~~~~~~~~~~~~~~~~~~~ \\
+\sum_{1} \chi k^{\lambda_1} \prod^{s-1}_{n=1} (e^{i \varphi} e^{\mu_n}_{+}
+e^{-i \varphi} e^{\mu_n}_{-})\oplus L_a +...+\chi
k^{\lambda_1}... \chi k^{\lambda_s}\oplus  L_{a} ,\nn
\eeqa
where the first term
$\prod^{s}_{n=1} (e^{i \varphi} e^{\mu_n}_{+}
+e^{-i \varphi} e^{\mu_n}_{-})$ represents the {\it helicity generators }
$(L^{+\cdot\cdot\cdot+}_{a},...,L^{-\cdot\cdot\cdot-}_{a})$, while their
helicity spectrum is described by the formula  (\ref{trasversalgenera1}).
The rest of the terms are purely
longitudinal and proportional to the increasing powers of momentum $k$.
The last formula also illustrates the
realisation of the transformation (\ref{isomorfism}), that is, the
helicity generators  $(L^{+\cdot\cdot\cdot+}_{a},...,L^{-\cdot\cdot\cdot-}_{a})$
are  defined modulo longitudinal terms proportional to
$k^{\lambda_1}...k^{\lambda_n}, n=1,...,s$.

The very fact that the representation of the generators $L_{a}^{\bot \lambda_1 ... \lambda_{s}}$ is transversal plays an important
role in the definition of the gauge field $\CA_{\mu}(x,e)$ in (\ref{gaugefield}).
Indeed, substituting the transversal representation (\ref{trasversalgenera2}) of the
generators $L_{a}^{\bot \lambda_1 ... \lambda_{s}}$ into the
expansion (\ref{gaugefield}) and collecting the terms in front of the helicity generators
$(L^{+\cdot\cdot\cdot+}_{a},...,L^{-\cdot\cdot\cdot-}_{a})$  we shall get
\beqa\label{gaugefield1}
{\cal A}_{\mu}(x,e)&=&  \sum_{s=0}^{\infty} {1\over s!} ~
 (\tilde{A}^{a}_{\mu \lambda_1 ... \lambda_{s}}~e^{\lambda_1}_{+}...e^{\lambda_s}_{+}
 \oplus L_{a} +...+
\tilde{A}^{a}_{\mu \lambda_1 ... \lambda_{s}} e^{\lambda_1}_{-}...e^{\lambda_s}_{-}\oplus L_{a} )     \nn\\
&=& \sum_{s=0}^{\infty} {1\over s!} ~
 (\tilde{A}^{a}_{\mu +\cdot\cdot\cdot+}~L^{+\cdot\cdot\cdot+}_{a} +...+
\tilde{A}^{a}_{\mu -\cdot\cdot\cdot-}~L^{-\cdot\cdot\cdot-}_{a} ),
\eeqa
where s is the number of negative indices.
This formula represents the projection  $\tilde{A}^{a}_{\mu \lambda_1 ... \lambda_{s}}$
 of the components  of
the non-Abelian tensor gauge field  $A^{a}_{\mu\lambda_1 ... \lambda_{s}} $ into the plane
transversal to the momentum.  The projection contains only positive definite space-like
components of the helicities \cite{Savvidy:2010vb,Savvidy:2010kw,Antoniadis:2011re}:
\be\label{tensorspectrum}
h~~=~~~\pm (s+1),~~ \begin{array}{c} \pm (s-1)\\ \pm (s-1) \end{array},~~
\begin{array}{c} \pm (s-3)\\ \pm (s-3) \end{array},~~....,
\ee
where the lower helicity states have double degeneracy. The analysis of the
kinetic terms of the Lagrangian and of the
corresponding equation of
motions, which will be considered in the next section,  confirms that indeed
the propagating degrees of freedom are described by helicities (\ref{tensorspectrum}).

In order to define the gauge invariant Lagrangian one should know the Killing
metric  of the algebra $L_G(\CP)$.
The explicit transversal representation of the $L_G(\CP)$ generators given above
(\ref{trasversalgenera}), (\ref{trasversalgenera11}) and (\ref{trasversalgenera2})
allows to calculate the corresponding Killing metric \cite{Savvidy:2010vb,Savvidy:2010kw,Savvidy:2013gsa}:
\beqa\label{killingform}
L_G:~~~~~~&&~~~~ \<L_{a}; L_{b} \rangle  =\delta_{ab}, \label{finale0}\\
\nn\\
L_{\CP}:~~~~~&& ~~~\<P^{\mu} ; P^{\nu }  \rangle ~=0\nn\\
&&~~~\<M_{\mu\nu} ; P_{\lambda  }  \rangle ~  =0\label{poincare0}\\
&& ~~~\<M^{\mu\nu} ; M^{\lambda \rho }  \rangle =\eta^{\mu\lambda } \eta^{\nu\rho}
-\eta^{\mu\rho} \eta^{\nu \lambda }\nn
\eeqa
\beqa
L_G(\CP):~~~~~&&~~~~~~\<P^{\mu};L_{a}^{\bot~ \lambda_1 ... \lambda_{s}}\rangle  =0,\nn\\
&&~~~~~~\<M^{\mu\nu};L_{a}^{\bot~ \lambda_1 ... \lambda_{s}}\rangle =0, \label{poincarecurrent0}\\
\nn\\
&&~~~~~~\<L_{a}; L^{\bot~\lambda_1}_{b} \rangle  =0,\nn\\
&&~~~~~~\<L^{\bot~\lambda_1}_{a}; L^{\bot~\lambda_2}_{b} \rangle  = \delta_{ab}~ \bar{\eta}^{\lambda_1 \lambda_2} ,\nn\\
&&~~~~~~\<L_{a}; L^{\bot~\lambda_1\lambda_2}_{b} \rangle  = \delta_{ab}~ \bar{\eta}^{\lambda_1 \lambda_2} ,\nn\\
&&~~~~~~\<L^{\bot~\lambda_1}_{a}; L^{\bot~\lambda_2 \lambda_3}_{b} \rangle  =0,\label{currentscalar0}\\
&&~~~~~~~~~~.....................\nn\\
&&~~~~~~\<L^{\bot~\lambda_1...\lambda_n}_{a}; L^{\bot~\lambda_{n+1}....\lambda_{2s+1}}_{b} \rangle  = 0,~~~~~~~~~s=0,1,2,3,...\nn\\
&&~~~~~~\<L^{\bot~\lambda_1...\lambda_n}_{a}; L^{\bot~\lambda_{n+1}....\lambda_{2s}}_{b} \rangle  =
\delta_{ab}~ s!~(\bar{\eta}^{\lambda_1 \lambda_2}  \bar{\eta}^{\lambda_3 \lambda_4}...
\bar{\eta}^{\lambda_{2s-1} \lambda_{2s}} +\textrm{perm}),\nn
\eeqa
where $\bar{\eta}^{\lambda_1\lambda_2}$ is the projector  into
the two-dimensional plane transversal to the momentum $k^\mu$ \cite{schwinger}:
\be\label{progector}
\bar{\eta}^{\lambda_1\lambda_2} =  { k^{\lambda_1}\bar{k}^{\lambda_2}
+\bar{k}^{\lambda_1}k^{\lambda_2} \over k \bar{k}}- \eta^{\lambda_1 \lambda_2} ,~~~~~~~
k_{\lambda_1 }\bar{\eta}^{\lambda_1\lambda_2}=
k_{\lambda_2 }\bar{\eta}^{\lambda_1\lambda_2}=0,
\ee
and $\bar{k}^{\mu}=\omega(1,0,0,-1)$.
It follows then that the transversality conditions (\ref{transversality}) are fulfilled:
\be\label{transversality1}
k_{\lambda_i} \<L^{\bot~\lambda_1...\lambda_n}_{a}; L^{\bot~\lambda_{n+1}....
\lambda_{2s}}_{b} \rangle =0,~~~~i=1,2,...2s.
\ee
The Killing metric on the internal $L_G$  and on the Poincar\'e  $L_{\CP}$
 subalgebras (\ref{finale0}), (\ref{poincare0}) are well known.
The important conclusion which follows from the above result is that the
Poincar\'e generators $P^{\mu}, M^{\mu\nu}$ are orthogonal to the  gauge  generators
$L_{a}^{ \lambda_1 ... \lambda_{s}}$ (\ref{poincarecurrent0}). The last formulas
(\ref{currentscalar0}) represent the Killing metric on the $L_{\CG}$ current
algebra (\ref{currentalge}),(\ref{extensionofpoincarealgebra}) and will be used
in the definition of the Lagrangian in the next section. It should be stressed that 
the metric (\ref{currentscalar0}) is defined modulo longitudinal terms. This is  
because under the "gauge" transformation of the generators (\ref{isomorfism}) 
the metric will receive  terms which are polynomial  in momentum. 
The provided metric (\ref{currentscalar0}) is written in a particular gauge. 
This peculiar property of the metric is mirrored in the definition of the Lagrangian 
which can be written in different gauges. The spectrum of the propagating 
modes does not depend on the gauges chosen, as one can get convinced 
by inspecting the expression (\ref{gaugefield1}).

Notice that
the reducible representation (\ref{represofextenpoincarealgebra}), without
any of  the constraints (\ref{constraint}), should also be considered,
as well as the representation in which only
the last constrain in (\ref{constraint}) is imposed. In that cases the transversality of the
representation (\ref{transversality1}) will be lost, but instead one arrives to the homogeneous
Killing metric in (\ref{currentscalar0}) $\bar{\eta}^{\lambda_1\lambda_2}\rightarrow \eta^{\lambda_1\lambda_2}$ and the longitudinal terms which can be gauged away.

{\it With this Killing metric in hands one
can define the Lagrangian of the theory}.

\section{\it The Lagrangian }

The gauge transformation of the field $\CA_{\mu}(x,e)$ is defined as \cite{yang,Savvidy:2005ki,Savvidy:2010vb}
\be\label{extendedgaugetransformation}
\CA^{'}_{\mu}(x,e) = U(\xi)  \CA_{\mu}(x,e) U^{-1}(\xi) -{i\over g}
\partial_{\mu}U(\xi) ~U^{-1}(\xi),
\ee
where the group parameter  $\xi(x,e)$
$$
U(\xi)=e^{i \xi(x,e)}
$$
has the decomposition  \cite{Savvidy:2005ki,Savvidy:2010vb}
$$
\xi(x,e)=  \sum_s {1\over s!}~\xi^{a}_{\lambda_1 ... \lambda_{s}}(x) ~~L_{a}e^{\lambda_{1}}...e^{\lambda_{s}}
$$
and $\xi^{a}_{\lambda_1 ... \lambda_{s}}(x)$ are totally symmetric gauge parameters.
Using the commutator of the covariant derivatives
$
\nabla^{ab}_{\mu} = (\partial_{\mu}-ig \CA_{\mu}(x,e))^{ab}
$
\be
[\nabla_{\mu}, \nabla_{\nu}]^{ab} = g f^{acb} \CG^{c}_{\mu\nu}~,
\ee
we can define the extended field strength tensor
\be\label{fieldstrengthgeneral}
\CG_{\mu\nu}(x,e) = \partial_{\mu} \CA_{\nu}(x,e) - \partial_{\nu} \CA_{\mu}(x,e) -
i g [ \CA_{\mu}(x,e)~\CA_{\nu}(x,e)],
\ee
which transforms homogeneously:
\be\label{fieldstrenghthtransformation}
\CG^{'}_{\mu\nu}(x,e)) = U(\xi)  \CG_{\mu\nu}(x,e) U^{-1}(\xi).
\ee
It is useful to have an explicit expression for the transformation law of the field components
 \cite{Savvidy:2005fi,Savvidy:2005zm,Savvidy:2005ki}:
\beqa\label{polygauge}
\delta A^{a}_{\mu} &=& ( \delta^{ab}\partial_{\mu}
+g f^{acb}A^{c}_{\mu})\xi^b ,~~~~~\\
\delta A^{a}_{\mu\nu} &=&  ( \delta^{ab}\partial_{\mu}
+  g f^{acb}A^{c}_{\mu})\xi^{b}_{\nu} + g f^{acb}A^{c}_{\mu\nu}\xi^{b},\nonumber\\
\delta A^{a}_{\mu\nu \lambda}& =&  ( \delta^{ab}\partial_{\mu}
+g f^{acb} A^{c}_{\mu})\xi^{b}_{\nu\lambda} +
g f^{acb}(  A^{c}_{\mu  \nu}\xi^{b}_{\lambda } +
A^{c}_{\mu \lambda }\xi^{b}_{ \nu}+
A^{c}_{\mu\nu\lambda}\xi^{b}),\nn\\
.........&.&............................ \nn
\eeqa
These extended gauge transformations generate a closed algebraic structure.
The  component   field strengths  tensors take the following form
\cite{Savvidy:2005fi,Savvidy:2005zm,Savvidy:2005ki}:
\beqa\label{fieldstrengthparticular}
G^{a}_{\mu\nu} &=&
\partial_{\mu} A^{a}_{\nu} - \partial_{\nu} A^{a}_{\mu} +
g f^{abc}~A^{b}_{\mu}~A^{c}_{\nu},\\
G^{a}_{\mu\nu,\lambda} &=&
\partial_{\mu} A^{a}_{\nu\lambda} - \partial_{\nu} A^{a}_{\mu\lambda} +
g f^{abc}(~A^{b}_{\mu}~A^{c}_{\nu\lambda} + A^{b}_{\mu\lambda}~A^{c}_{\nu} ~),\nn\\
G^{a}_{\mu\nu,\lambda\rho} &=&
\partial_{\mu} A^{a}_{\nu\lambda\rho} - \partial_{\nu} A^{a}_{\mu\lambda\rho} +
g f^{abc}(~A^{b}_{\mu}~A^{c}_{\nu\lambda\rho} +
 A^{b}_{\mu\lambda}~A^{c}_{\nu\rho}+A^{b}_{\mu\rho}~A^{c}_{\nu\lambda}
 + A^{b}_{\mu\lambda\rho}~A^{c}_{\nu} ~),\nn\\
 ......&.&............................................\nn
\eeqa
and transform homogeneously with respect to the
  transformations (\ref{polygauge}):
\beqa\label{fieldstrenghparticulartransformation}
\delta G^{a}_{\mu\nu}&=& g f^{abc} G^{b}_{\mu\nu} \xi^c  ,\\
\delta G^{a}_{\mu\nu,\lambda} &=& g f^{abc} (~G^{b}_{\mu\nu,\lambda} \xi^c
+ G^{b}_{\mu\nu} \xi^{c}_{\lambda}~),\nonumber\\
\delta G^{a}_{\mu\nu,\lambda\rho} &=& g f^{abc}
(~G^{b}_{\mu\nu,\lambda\rho} \xi^c
+ G^{b}_{\mu\nu,\lambda} \xi^{c}_{\rho} +
G^{b}_{\mu\nu,\rho} \xi^{c}_{\lambda} +
G^{b}_{\mu\nu} \xi^{c}_{\lambda\rho}~),\nn\\
......&.&..........................\nn
\eeqa
The field strength tensors are
antisymmetric in their first two indices and are totally symmetric with respect to the
rest of the indices. The symmetry properties of the field strength  $G^{a}_{\mu\nu,\lambda_1 ... \lambda_s}$ remain invariant in the course of these transformations.

\begin{figure}
\begin{center}
\includegraphics[width=3cm]{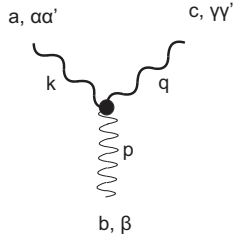}
\caption{The interaction vertex for the vector gauge boson V and two
tensor gauge bosons T - the VTT vertex -
$\CV^{abc}_{\alpha\acute{\alpha}\beta\gamma\acute{\gamma}}(k,p,q)$
 in non-Abelian tensor gauge
field theory \cite{Savvidy:2005ki}.
Vector gauge bosons are conventionally drawn
as thin wave lines, tensor gauge bosons are thick wave lines.
The Lorentz indices $\alpha\acute{\alpha}$ and momentum $k$ belong to the
first tensor gauge boson, the $\gamma\acute{\gamma}$ and momentum $q$
belong to the second tensor gauge boson, and Lorentz index $\beta$  and
momentum $p$ belong to the vector gauge boson. }
\label{fig1}
\end{center}
\end{figure}

\begin{figure}
\begin{center}
\includegraphics[width=3cm]{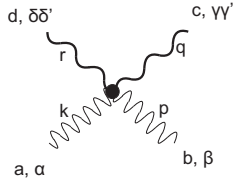}
\caption{The quartic vertex with two vector gauge bosons and two
tensor gauge bosons - the VVTT vertex -
${{\cal V}}^{abcd}_{\alpha\beta\gamma\acute{\gamma}\delta\acute{\delta}}(k,p,q,r)$
in non-Abelian tensor gauge
field theory \cite{Savvidy:2005ki}.
Vector gauge bosons are conventionally drawn
as thin wave lines, tensor gauge bosons are thick wave lines.
The Lorentz indices $\gamma\acute{\gamma}$ and momentum $q$ belong to the
first tensor gauge boson, $\delta\acute{\delta}$ and momentum $r$ belong to the
second tensor gauge boson, the index $\alpha$  and momentum $k$
belong to the first vector gauge boson and Lorentz index $\beta$  and
momentum $p$ belong to the second vector gauge boson.
}
\label{fig2}
\end{center}
\end{figure}

The first gauge invariant density is given by the expression \cite{Savvidy:2005fi,Savvidy:2005zm,Savvidy:2005ki}
\be\label{lagrangdensity}
{{\cal L}}(x)= \<{{\cal L}}(x,e)\rangle =  -{1\over 4} \<\CG^{a}_{\mu\nu}(x,e)\CG^{a \mu\nu}(x,e)\rangle,
\ee
where the trace of the generators is given in (\ref{currentscalar0}).
One can get convinced that the variation of the (\ref{lagrangdensity}) with respect to the
gauge transformations (\ref{extendedgaugetransformation})
and  (\ref{fieldstrenghthtransformation}) vanishes:
$$
\delta {{\cal L}}(x,e) = -{1\over 2}\CG^{a}_{\mu\nu}(x,e)~ g f^{abc}~
\CG^{b \mu\nu}(x,e) ~\xi^{c}(x,e) =0.
$$
The invariant density (\ref{lagrangdensity}) allows to extract
{\it gauge invariant, totally symmetric, tensor densities
$\CL_{\lambda_1 ... \lambda_{s}}(x)$} by
using expansion with respect to the vector variable $e^{\lambda}$:
\be
\CL(x,e) = \sum^{\infty}_{s=0}~{1\over s!}
\CL_{\lambda_1 ... \lambda_{s}}(x) ~ e^{\lambda_1}...e^{\lambda_s} .
\ee
In particular, the expansion term which is quadratic in powers of $e^{\lambda}$ is
\be
\CL_{\lambda_1\lambda_2} = -{1\over 4}G^{a}_{\mu\nu,\lambda_1}G^{a}_{\mu\nu,\lambda_2}
-{1\over 4}G^{a}_{\mu\nu}G^{a}_{\mu\nu,\lambda_1\lambda_2}.
\ee
The gauge invariant density thus can be represented in the following form
\cite{Savvidy:2005fi,Savvidy:2005zm,Savvidy:2005ki}:
\be
\CL(x) = \<\CL(x,e)\rangle = \sum^{\infty}_{s=0}~{1\over s!}
\CL_{\lambda_1 ... \lambda_{s}}(x) ~ \<e^{\lambda_1}...e^{\lambda_s}\rangle
\ee
and the density for the lower-rank  tensor fields is
$$
{{\cal L}}_2 =-{1\over 4}G^{a}_{\mu\nu,\lambda}G^{a}_{ \mu\nu,\lambda}
-{1\over 4}G^{a}_{\mu\nu}G^{a}_{ \mu\nu,\lambda \lambda}.
$$
Let us consider the second gauge invariant   density of the form \cite{Savvidy:2005fi,Savvidy:2005zm,Savvidy:2005ki}
\be\label{secondseries}
{\cal L}^{'}(x)=\<{\cal L}^{'}(x,e)\rangle = {1\over 4}
\<\CG^{a}_{\mu\rho_1}(x,e) e^{\rho_1} ~\CG^{a \mu}_{~~~~\rho_2}(x,e) e^{\rho_2}\rangle^{'}.
\ee
It is gauge invariant because its variation is also equal to zero:
\beqa
\delta {{\cal L}}^{'}(x,e) ={1\over 4}g f^{acb}~
\CG^{c}_{\mu\rho_1}(x,e)e^{\rho_1} ~\xi^{b}(x,e)
\CG^{a\mu}_{~~~~\rho_2}(x,e)e^{\rho_2}+\nn
\\
+{1\over 4}\CG^{a}_{\mu\rho_1}(x,e)e^{\rho_1}~ g f^{acb}~
\CG^{c\mu}_{~~~~\rho_2}(x,e) e^{\rho_2}~\xi^{b}(x,e) =0.
\eeqa
The Lagrangian density (\ref{secondseries}) generates the
second series of {\it gauge invariant tensor densities
$(\CL^{'}_{\rho_1\rho_2})_{\lambda_1 ... \lambda_{s}}(x)$}
when we expand it in powers of the vector variable $e^{\lambda}$:
\be\label{secondseriesdensities}
{\cal L}^{'}(x)=\<{\cal L}^{'}(x,e)\rangle= \sum^{\infty}_{s=0}~{1\over s!}
(\CL^{'}_{\rho_1\rho_2})_{\lambda_1 ... \lambda_{s}}(x) ~\<e^{\rho_1}e^{\rho_2}
e^{\lambda_1}...e^{\lambda_s} \rangle^{'}.
\ee
The term quartic in variable $e^{\lambda}$ after contraction of the vector
variables takes the following form:
\be\label{actiontwoprime}
{{\cal L}}^{'}_2 =  {1\over 4}G^{a}_{\mu\nu,\lambda}G^{a}_{\mu\lambda,\nu}
+{1\over 4}G^{a}_{\mu\nu,\nu}G^{a}_{\mu\lambda,\lambda}
+{1\over 2} G^{a}_{\mu\nu}G^{a}_{\mu\lambda,\nu\lambda} .
\ee
One can get convinced that it is gauge invariant under the
transformation (\ref{polygauge}) and (\ref{fieldstrenghparticulartransformation}).
The total Lagrangian density is a sum of two invariants (\ref{lagrangdensity})
and (\ref{secondseries}):
\be\label{total}
L = \CL + \CL^{'}=-{1\over 4} \<\CG^{a}_{\mu\nu}(x,e)\CG^{a \mu\nu}(x,e)\rangle +
{1\over 4}
\<\CG^{a}_{\mu\rho_1}(x,e) e^{\rho_1} ~\CG^{a \mu}_{~~~~\rho_2}(x,e) e^{\rho_2}\rangle^{'}.
\ee
The Lagrangian for the lower-rank tensor gauge fields has the following form:
\beqa\label{totalactiontwo}
{{\cal L}}=  {{\cal L}}_1 +  {{\cal L}}_2 +  {{\cal L}}^{'}_2 +...=
&-&{1\over 4}G^{a}_{\mu\nu}G^{a}_{\mu\nu}\\
&-&{1\over 4}G^{a}_{\mu\nu,\lambda}G^{a}_{\mu\nu,\lambda}
-{1\over 4}G^{a}_{\mu\nu}G^{a}_{\mu\nu,\lambda\lambda}\nn\\
&+&{1\over 4}G^{a}_{\mu\nu,\lambda}G^{a}_{\mu\lambda,\nu}
+{1\over 4}G^{a}_{\mu\nu,\nu}G^{a}_{\mu\lambda,\lambda}
+{1\over 2}G^{a}_{\mu\nu}G^{a}_{\mu\lambda,\nu\lambda} +...\nn
\eeqa
The above Lagrangian defines the kinetic operators for the rank-1 $A^a_{\mu}$
and rank-2  $A^{a}_{\mu\lambda_1}$ fields,
as well as trilinear  and quartic  interactions
 with the {\it dimensionless coupling constant g} (see Fig.\ref{fig1}-\ref{fig2}) .

As we found in
\cite{Savvidy:2005fi,Savvidy:2005zm,Savvidy:2005ki},
the corresponding free field equations coincide with the equations introduces in
the classical works \cite{fierz,fierzpauli,minkowski} and
describe the propagation of the {\it helicity-two and zero $h = \pm 2,0 $
 massless charged
tensor gauge bosons}, and there are no propagating negative norm states.
This is in agreement with the spectrum presented in (\ref{tensorspectrum}).
The next term in expansion of the Lagrangian density has the following
form \cite{Savvidy:2009wr,Savvidy:2009zz}:
\beqa\label{thirdranktensorlagrangian}
{\cal L}_3 +  {\cal L}^{'}_3
=&-&{1\over 4}G^{a}_{\mu\nu,\lambda\rho}G^{a}_{\mu\nu,\lambda\rho}
-{1\over 8}G^{a}_{\mu\nu ,\lambda\lambda}G^{a}_{\mu\nu ,\rho\rho}
-{1\over 2}G^{a}_{\mu\nu,\lambda}  G^{a}_{\mu\nu ,\lambda \rho\rho}
-{1\over 8}G^{a}_{\mu\nu}  G^{a}_{\mu\nu ,\lambda \lambda\rho\rho}+ \nn\\
&+&{1\over 3}G^{a}_{\mu\nu,\lambda\rho}G^{a}_{\mu\lambda,\nu\rho}+
{1\over 3} G^{a}_{\mu\nu,\nu\lambda}G^{a}_{\mu\rho,\rho\lambda}+
{1\over 3}G^{a}_{\mu\nu,\nu\lambda}G^{a}_{\mu\lambda,\rho\rho}+\\
&+&{1\over 3}G^{a}_{\mu\nu,\lambda}G^{a}_{\mu\lambda,\nu\rho\rho}
+{2\over 3}G^{a}_{\mu\nu,\lambda}G^{a}_{\mu\rho,\nu\lambda\rho}
+{1\over 3}G^{a}_{\mu\nu,\nu}G^{a}_{\mu\lambda,\lambda\rho\rho}
+{1\over 3}G^{a}_{\mu\nu}G^{a}_{\mu\lambda,\nu\lambda\rho\rho}\nn
\eeqa
and the corresponding free field equations for the tensor
gauge field $A_{\mu\lambda_1\lambda_2}$
in four-dimensional space-time
describe the {\it propagation of helicity-three and one
 $h= \pm 3, \pm 1,\pm 1 $ massless charged gauge bosons} in
agrement with the spectrum (\ref{tensorspectrum}).
There are no propagating negative norm states.   The comparison
of these equations with the Schwinger-Fronsdal equations
\cite{schwinger,singh,fronsdal,Weinberg:1964cn}
can be found in \cite{Guttenberg:2008qe}.

Considering the free field equation for the general
rank-(s+1) tensor gauge field one can find that
the quadratic part of the Lagrangian has the following form \cite{Savvidy:2009zz}:
\beqa\label{totalfreelagrangianranks}
{{\cal L}}_{s+1} +   { {\cal L}}^{'}_{s+1} ~\vert_{quadratic}=
 {1 \over 2} A^{a}_{\alpha \lambda_{1}...\lambda_{s} }
\CH^{\alpha \lambda_{1}...\lambda_{s} ~ \gamma \lambda_{s+1}...\lambda_{2s}  }
A^{a}_{\gamma \lambda_{s+1}...\lambda_{2s}  }
\eeqa
and is invariant with respect to the group of gauge transformations
\be\label{fullgroupofextendedtransformations}
\delta A^{a}_{\alpha \lambda_{1}...\lambda_{s}}
=\partial_{\alpha} \xi^{a}_{\lambda_{1}...\lambda_{s}},~~~~~~~~~
\tilde{\delta} A^{a}_{\alpha \lambda_{1}...\lambda_{s}} =
\partial_{\lambda_{1}} \zeta^{a}_{\lambda_{2}...\lambda_{s}\alpha }+...
+\partial_{\lambda_{s}} \zeta^{a}_{\lambda_{1}...\lambda_{s-1}\alpha },
\ee
which should fulfil the following constraints:
\beqa
\begin{array}{cc}
 \partial_{\rho}\zeta^{a}_{\rho\lambda_1...\lambda_{s-1}}-
{1 \over s-2}(\partial_{\lambda_1} \zeta^{a}_{\lambda_2...\lambda_{s-1} \rho\rho}+
...+ \partial_{\lambda_{s-1}} \zeta^{a}_{\lambda_1...\lambda_{s-2} \rho\rho})=0, \\
\\
 \partial_{\lambda_1} \zeta^{a}_{\lambda_2...\lambda_{s-1} \rho\rho}-
\partial_{\lambda_2} \zeta^{a}_{\lambda_1...\lambda_{s-1} \rho\rho}=0.
\end{array}
\eeqa
In momentum representation the kinetic operator has the following general form:
\beqa\label{quadraticformranks}
\CH_{\alpha \lambda_{1}...\lambda_{s} ~ \gamma\lambda_{s+1}...\lambda_{2s}  }=
&+&{1 \over s!}~
(\sum_{p} \eta_{\lambda_{i_1} \lambda_{i_2}} .......
\eta_{\lambda_{i_{2s-1}} \lambda_{i_{2s}}})~
(-k^2 \eta_{\alpha\gamma} + k_{\alpha}k_{\gamma})\nn
\\
 &+& {1 \over (s+1)!}  (\sum_{P} \eta_{ \alpha \lambda_{i_1}}~
 \eta_{\lambda_{i_2 } \lambda_{ i_3} }.......
 \eta_{ \lambda_{ i_{2s-2} } \lambda_{i_{2s-1}} }~
\eta_{ \gamma  \lambda_{i_{2s}}})~k^2
\nn\\
             &-&{ 1\over  (s+1)!}  ~
(\sum_{P } \eta_{\rho \lambda_{i_1}}~
\eta_{\lambda_{i_2} \lambda_{i_3}} .......
                 \eta_{ \lambda_{ i_{2s-2} } \lambda_{i_{2s-1}} }~
\eta_{ \gamma  \lambda_{i_{2s}}})
                 ~
 k_{\alpha} k_{\rho}
 \\
 & -&{ 1\over  (s+1)!}  ~
(\sum_{P} \eta_{\rho \lambda_{i_1}}~
\eta_{\lambda_{i_2} \lambda_{i_3}} .......
                 \eta_{ \lambda_{ i_{2s-2} } \lambda_{i_{2s-1}} }~
\eta_{ \alpha  \lambda_{i_{2s}}}) ~
 k_{\rho} k_{\gamma}
\nn\\
   &+ &{ 1\over (s+1)! } ~\eta_{\alpha\gamma}~
   (\sum_{P}
   \eta_{\rho \lambda_{i_1}}~
\eta_{\lambda_{i_2} \lambda_{i_3}} .......
                 \eta_{ \lambda_{ i_{2s-2} } \lambda_{i_{2s-1}} }~
\eta_{ \sigma \lambda_{i_{2s}}}) ~
 k_{\rho } k_{\sigma}
,\nn
\eeqa
where the sum $\sum_P$ runs over all non-equal permutations of
$\lambda_i~'s$.  The solution of the free field equation for the rank-(s+1)
field \cite{Savvidy:2009zz}
\be\label{generalrankequation}
\CH^{\alpha \lambda_{1}...\lambda_{s} ~ \gamma \lambda_{s+1}...\lambda_{2s}  }~
A_{\gamma \lambda_{s+1}...\lambda_{2s}  }=0.
\ee
describes the propagation of the helicities:
\be\label{spectrum}
h= \pm (s+1),~~ \begin{array}{c} \pm (s-1)\\ \pm (s-1) \end{array},~~
\begin{array}{c} \pm (s-3)\\ \pm (s-3) \end{array},~~....
\ee
It is convenient to represent the spectrum (\ref{spectrum}) of tensor gauge bosons 
in the form which combines the helicity spectrum of all bosons. It is unbounded
and  has the following form \cite{Savvidy:2010vb}:
\beqa\label{spectrum1}
&&\pm 1\nn\\
&&\pm 2, ~~~~~0\nn\\
&&\pm 3,~~\pm 1,~~\pm 1\nn\\
&&\pm 4,~~\pm 2,~~\pm 2,~~~~0\nn\\
&&\pm 5,~~\pm 3,~~\pm 3,~~\pm 1,~~\pm 1  \nn\\
&&\pm 6,~~\pm 4,~~\pm 4,~~\pm 2,~~\pm 2,~~~~0\nn\\
&&....................................................
\eeqa
In summary, we defined the composite gauge field (\ref{gaugefield}) which takes a value in
the transversal representation (\ref{trasversalgenera}), (\ref{trasversalgenera11}),
(\ref{trasversalgenera2}) of the extended Poincar\'e algebra $L_G(\CP)$. We
constructed the invariant Lagrangian (\ref{total}), (\ref{lagrangdensity}), (\ref{secondseries})
which contains infinity many tensor gauge fields (\ref{tensorfields}) and found their
helicity content (\ref{spectrum}), (\ref{spectrum1}).

The theory has unexpected symmetry with respect to the duality transformation of the gauge 
fields \cite{Barrett:2007nn,Guttenberg:2008zn}. The complementary gauge transformation   $\tilde{\delta} $ is defined as:
\beqa\label{doublepolygaugesymmetric}
\tilde{\delta}  A^{a}_{\mu} &=& ( \delta^{ab}\partial_{\mu}
+g f^{acb}A^{c}_{\mu})\eta^b ,\nn\\
\tilde{\delta}  A^{a}_{\mu\lambda_1} &=&  ( \delta^{ab}\partial_{\lambda_1}
+  g f^{acb}A^{c}_{\lambda_1})\eta^{b}_{\mu} + g f^{acb}A^{c}_{\mu\lambda_1}\eta^{b},\\
\tilde{\delta}  A^{a}_{\mu\lambda_1\lambda_2} &=& ( \delta^{ab}\partial_{\lambda_1}
+g f^{acb} A^{c}_{\lambda_1})\eta^{b}_{\mu\lambda_2} +( \delta^{ab}\partial_{\lambda_2}
+g f^{acb} A^{c}_{\lambda_2})\eta^{b}_{\mu\lambda_1} +\nn\\
&~&
+g f^{acb}(  A^{c}_{\mu  \lambda_1}\eta^{b}_{\lambda_2 }+
A^{c}_{\mu \lambda_2 }\eta^{b}_{\lambda_1}+
A^{c}_{\lambda_1\lambda_2}\eta^{b}_{\mu} +
A^{c}_{\lambda_2 \lambda_1}\eta^{b}_{\mu} +A^{c}_{\mu\lambda_1\lambda_2}\eta^{b}),\nn\\
.........&.&............................\nn
\eeqa
The transformations $\delta$ in (\ref{polygauge}) and $\tilde{\delta} $ in 
(\ref{doublepolygaugesymmetric}) do not
coincide and are {\it complementary} to each other in the following sense:
in $\delta$
the derivatives of the gauge parameters $\{  \xi \}$ are  over the first
index $\mu$, while in $\tilde{\delta} $ the derivatives of the
gauge parameters $\{ \eta \}$ are over the rest of the totally symmetric
indices $\lambda_1 ... \lambda_{s}$.
 One can  construct the  
new field strength tensors
$\tilde{G}^{a}_{\mu\nu,\lambda_1 ... \lambda_s}$
which are transforming  homogeneously 
with respect to the $\tilde{\delta}$ transformations and then 
to construct the corresponding gauge invariant Lagrangian
$\tilde{L}(A)$ \cite{Barrett:2007nn,Guttenberg:2008zn}. The relation between these two Lagrangians 
was found in the form of duality transformation \cite{Barrett:2007nn,Guttenberg:2008zn}:
\beqa\label{dualtransformationincompo}
\begin{array}{ll}
\tilde{A}_{\mu\lambda_1} =  A_{\lambda_1\mu}   ,  \\
\tilde{A}_{\mu\lambda_1\lambda_2} =
{1\over 2}(A_{\lambda_1\mu\lambda_2} + A_{\lambda_2\mu\lambda_1})
-{1\over 2} A_{\mu\lambda_1\lambda_2}, \\
\tilde{A}_{\mu\lambda_1\lambda_2\lambda_3} =
{1\over 3}(A_{\lambda_1\mu\lambda_2\lambda_3}
+ A_{\lambda_2\mu\lambda_1\lambda_3}
+ A_{\lambda_3\mu\lambda_1\lambda_2})
-{2\over 3} A_{\mu\lambda_1\lambda_2 \lambda_3 }, \\
.........................................
 \end{array}
\eeqa
which maps the Lagrangian $L(\tilde{A})$
into the Lagrangian $\tilde{L}(A)$. This takes place
because  
$
G_{\mu\nu,\lambda_1 ... \lambda_s}(\tilde{A})=
\tilde{G}_{\mu\nu,\lambda_1 ... \lambda_s}(A) 
$
and therefore
$
L(\tilde{A})~= ~\tilde{L}(A) .
$

{\it The Lagrangian (\ref{total}) defines not only a free propagation
of tensor gauge bosons, but also their
interactions. The interaction diagrams for the lower-rank bosons are
presented on Fig.\ref{fig1}-\ref{fig2}. The high-rank bosons also interact
through the triple and quartic interaction vertices. It is therefore important
to  calculate and study
the scattering amplitudes, the quantum loop corrections and their high energy
behaviour. By using the diagram technique it is possible to
calculate the scattering amplitude, but the difficulties
lie in the evaluation and contraction of high-rank tensors structures appearing in the diagram
approach. In the next section we shall use alternative approach based on spinor representation
of amplitudes developed recently in \cite{Berends:1981rb,Kleiss:1985yh,Xu:1986xb,
Gunion:1985vca,Dixon:1996wi,Parke:1986gb,Berends:1987me,
Witten:2003nn,Cachazo:2004by,Cachazo:2004dr,Britto:2004ap,Britto:2005fq,
Benincasa:2007xk,Cachazo:2004kj,Georgiou:2004by,
ArkaniHamed:2008yf,Berends:1988zn,Mangano:1987kp}.}

\section{\it Scattering Amplitudes and Splitting Functions}

A scattering amplitude for the
massless particles of momenta $p_i$ and polarisation tensors $\varepsilon_i$ ~$(i=1,...,n)$,
which are described by irreducible massless representations of the Poincar\'e group,
can be represented in the following form:
$$
M_n = M_n(p_1,\varepsilon_1;~p_2,\varepsilon_2;~...;~p_n,\varepsilon_n).
$$
It is more convenient to represent  the momenta $p_i$ and polarisation tensors $\varepsilon_i$ in terms of spinors. In that case
the scattering amplitude   $M_n$ can  be considered
as a function of spinors $\lambda_i$, $\tilde{\lambda}_i$ and helicities $h_i$
\cite{Berends:1981rb,Kleiss:1985yh,Xu:1986xb,
Gunion:1985vca,Dixon:1996wi,Parke:1986gb,Berends:1987me,
Witten:2003nn,Cachazo:2004by,Cachazo:2004dr,Britto:2004ap,Britto:2005fq,
Benincasa:2007xk,Cachazo:2004kj,Georgiou:2004by,
ArkaniHamed:2008yf,Berends:1988zn,Mangano:1987kp}:
\be\label{smatrix}
M_n=M_n(\lambda_1,\tilde{\lambda}_1,h_1;~...;~\lambda_n,\tilde{\lambda}_n,h_n)~.
\ee
The advantage of the spinor representation is that introducing a complex deformation of
the particles momenta
one can derive a general form for the three-particle interaction vertices
\cite{Bengtsson:1983pd,Bengtsson:1983pg,Benincasa:2007xk,Georgiou:2010mf,Savvidy:2014uva}:
$$
M_3(1^{h_1} ,2^{h_2},3^{h_3} ).
$$
The dimensionality of the three-point  vertex  $M_3(1^{h_1} ,2^{h_2},3^{h_3} )$ is
$$
[mass]^{D=\pm(h_1+h_2+h_3)}.
$$
In the generalised Yang-Mills theory \cite{Savvidy:2005fi,Savvidy:2005zm,Savvidy:2005ki,Savvidy:2010vb},  which we  described in the previous sections,
all interaction vertices
between high-spin particles have {\it dimensionless coupling constants},
which means that the helicities of the interacting particles in the vertex are
constrained  by the relation
$$
D=\pm(h_1+h_2+h_3)= 1~.
$$
Therefore the interaction vertex between
massless tensor-bosons, the TTT-vertex,   has the following
general form \cite{Georgiou:2010mf,Savvidy:2014uva}:
\beqa\label{dimensionone1}
M_3 &=& g f^{abc} <1,2>^{-2h_1 -2h_2 -1} <2,3>^{2h_1 +1} <3,1>^{2h_2 +1},~~~~h_3= -1 - h_1 -h_2, \nn\\
M_3 &=& g f^{abc} [1,2]^{2h_1 +2h_2 -1} [2,3]^{-2h_1 +1} [3,1]^{-2h_2 +1},~~~~~h_3= 1 - h_1 -h_2,
\eeqa
where $f^{abc}$ are the structure constants of the internal gauge group G.
In particular, considering the interaction between a boson of helicity $h_1 = \pm 1$ and
a tensor-boson of helicity $h_2 = \pm s$, the VTT-vertex,  one can find from  (\ref{dimensionone1})  that
\be\label{vertecies}
h_3 = \pm \vert s-2 \vert,~ \pm s,~
 \pm \vert s+2 \vert~
\ee
and the corresponding vector-tensor-tensor  interaction vertices VTT have
the following form:
\beqa\label{1ssvertex}
M^{a_1a_2a_3}_3(1^{-s} ,2^{-1} ,3^{+s} )&=& g ~f^{a_1 a_2 a_3} {<1,2>^{4} \over <1,2> <2,3> <3,1>}
\left({<1,2>  \over  <2,3> }\right)^{2s-2},\nn\\
M^{a_1a_2a_3}_3(1^{-s} ,2^{+1},3^{s-2} )&=& g ~f^{a_1 a_2 a_3} {<1,3>^{4} \over <1,2> <2,3> <3,1>}
\left({<1,2>  \over  <2,3> }\right)^{2s-2}.
\eeqa
These are the vertices which reduce to the standard triple YM  vertex  when $s=1$.
Using these vertices  one can compute the scattering amplitudes of vector and  tensor bosons.
The colour-ordered
scattering amplitudes involving two tensor-bosons of helicities $h =\pm s$,
one negative helicity vector-boson
and $(n-3)$ vector-bosons  of positive helicity were found  in  \cite{Georgiou:2010mf}:
\be\label{gs}
\hat{M}_n(1^+,..i^-,...k^{+s},..j^{-s},..n^+)=i g^{n-2} (2\pi)^4 \delta^{(4)}(P^{a\dot{b}})
 \frac{<ij>^4}{\prod_{l=1}^{n} <l l+1>} \Big( \frac{<ij>}{<ik>}\Big)^{2s-2},
\ee
where $n$ is the total number of particles and the dots stand for any number of
positive helicity vector-bosons, $i$ is the position of the negative-helicity vector,
while $k$ and $j$ are the positions of the tensors with helicities $+s$ and $-s$ respectively.
The expression \eqref{gs}  reduces to the
famous Parke-Taylor formula \cite{Parke:1986gb} when $s=1$.
In particular, the five-particle amplitude takes the following form:
\beqa\label{fivepointamplitude}
\hat{M}_5(1^+,2^-,3^{+},4^{+s},5^{-s})=  i g^{3} (2\pi)^4 \delta^{(4)}(P^{a\dot{b}})
\frac{<25>^4}{\prod_{i=1}^{5}
<i i+1>}  (\frac{<25> }{<24>})^{2s-2},
\eeqa
where
$
P^{a\dot{b}} = \sum^n_{m=1} \lambda^a_m \tilde{\lambda}^{\dot{b}}_m
$ is the total momentum. Notice that the scattering amplitudes (\ref{gs}) and (\ref{fivepointamplitude})
have large validity area:  in the limit $s \rightarrow 1/2$ they
reduce to the tree level gluon scattering amplitudes into a quark pair and
into a pair of scalars as $s\rightarrow 0$.

The scattering amplitudes (\ref{gs}) and (\ref{fivepointamplitude})
can be used to extract splitting amplitudes
of vector and tensor bosons \cite{Antoniadis:2011rr}. The collinear behaviour
of the tree amplitudes has the following factorised  form
\cite{Dixon:1996wi,Parke:1986gb,Berends:1987me,Berends:1988zn,Mangano:1987kp}:
\be\label{factorization}
M^{tree}_n(...,a^{\lambda_a},b^{\lambda_b},...)~~  {a \parallel b \over \rightarrow}  ~~  \sum_{\lambda=\pm 1}
Split^{ tree }_{-\lambda}(a^{\lambda_a},b^{\lambda_b})~ \times ~M^{tree}_{n-1}(...,P^{\lambda} ,...),
\ee
where $Split^{tree}_{-\lambda}(a^{\lambda_a},b^{\lambda_b})$ denotes the splitting amplitude and the
intermediate state $P$ has momentum $k_P=k_a +k_b$ and helicity $\lambda$.
Considering the amplitude \eqref{fivepointamplitude} in the limit when the
particles 4 and 5 become collinear, $k_4  \parallel k_5$, that is,
$k_4 = z k_P,~k_5 = (1-z) k_P$,  $k^2_P \rightarrow 0$ and $z$ describes the
longitudinal momentum sharing, one can deduce that the corresponding behaviour of spinors is
$
\lambda_4 = \sqrt{z} \lambda_P,~~~\lambda_5 = \sqrt{1-z} \lambda_P,
$
and  that the amplitude \eqref{fivepointamplitude} takes the following factorisation
form \cite{Antoniadis:2011rr}:
\beqa\label{factor}
M_5(1^+,2^-,3^{+},4^{+s},5^{-s})
 = A_4(1^+,2^-,3^{+},P^-) \times ~ Split_+(a^{+s},b^{-s}),
\eeqa
where
\be\label{spliting1}
Split_+(a^{+s},b^{-s}) = \left(\frac{1-z}{z } \right)^{ s-1}  \frac{(1-z)^2}{\sqrt{z(1-z)}}
\frac{1}{ <a, b>}.
\ee
In a similar way one can deduce that
\be\label{spliting2}
Split_+(a^{-s},b^{+s}) = \left(\frac{z }{1-z} \right)^{ s-1}  \frac{z^2}{\sqrt{z(1-z)}}
\frac{1}{ <a, b>}.
\ee
Considering different collinear limits $k_1  \parallel k_5$ and $k_3  \parallel k_4$
one can get \cite{Antoniadis:2011rr}
\be\label{spliting3}
Split_{+s}(a^{+},b^{-s}) =  \frac{(1-z)^{s+1}}{\sqrt{z(1-z)}}
\frac{1}{ <a, b>},~~~
Split_{+s}(a^{-s},b^{+}) =  \frac{z^{s+1}}{\sqrt{z(1-z)}}
\frac{1}{ <a, b>}
\ee
and
\be\label{spliting4}
Split_{-s}(a^{+s},b^{+}) =  \frac{z^{-s+1}}{\sqrt{z(1-z)}}
\frac{1}{ <a, b>}, ~~~
Split_{-s}(a^{+},b^{+s}) =    \frac{(1-z)^{-s+1}}{\sqrt{z(1-z)}}
\frac{1}{ <a, b>}.
\ee
The set of splitting amplitudes (\ref{spliting1})-(\ref{spliting4})
 $V\rightarrow TT$, $T \rightarrow VT$ and
$T \rightarrow TV$ reduces to the full set of gluon splitting amplitudes
\cite{Dixon:1996wi,Parke:1986gb,Berends:1987me,Berends:1988zn,Mangano:1987kp}
when $s=1$.

Since the collinear limits of the scattering amplitudes
are responsible for parton evolution  \cite{Altarelli:1977zs}
we can extract from the above expressions the
Altarelli-Parisi splitting probabilities for
tensor-bosons. Indeed, the residue of the collinear
pole in the square (of the factorised
amplitude (\ref{factorization})) gives Altarelli-Parisi splitting probability $P(z)$:
\be\label{AltarelliParisi}
P(z)= C_2(G) \sum_{h_P , h_a, h_b} \vert Split_{-h_P}(a^{h_a},b^{h_b}) \vert^2 ~ s_{ab},
\ee
where $s_{ab}=2 k_a \cdot k_b= <a,b>[a,b]$.
The invariant operator $C_2$ for the representation R is defined by the equation
$ t^a t^a  = C_2(R)~ 1 $ and $tr(t^a t^b) = T(R) \delta^{ab}$.
Substituting the splitting amplitudes (\ref{spliting1})-(\ref{spliting4})
into (\ref{AltarelliParisi}) we are getting
\beqa\label{setoftensorgluon}
P_{TV}(z) &=&  C_2(G)\left[ {z^4 \over z(1-z)}\left( {z\over 1-z} \right)^{2s-2}
+{(1-z)^4 \over z(1-z)} \left( {1-z\over  z}\right)^{2s-2} \right],\nn\\
P_{VT}(z) &=&  C_2(G)\left[ {1\over z(1-z)}\left( {1\over 1-z} \right)^{2s-2}
+{(1-z)^4 \over z(1-z)} (1-z)^{2s-2} \right],\\
P_{TT}(z) &=&  C_2(G)\left[ {z^4 \over z(1-z)} z^{2s-2}
+{1 \over z(1-z)} \left( {1\over  z}\right)^{2s-2} \right]. \nn
\eeqa
\begin{figure}
\includegraphics[width=6cm]{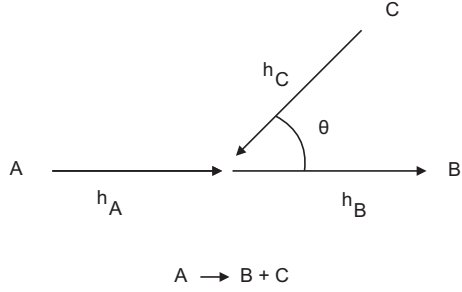}
\centering
\caption{The decay of a gluon of helicity $h_A$ into
the tensorgluons of helicities  $h_B$ and $h_C$. The arrows
show the directions of the helicities.
The corresponding splitting probability is defined as $P_{BA}$.  }
\label{fig3}
\end{figure}
The momentum conservation in the vertices clearly fulfils because these
functions satisfy the relations
\be
P_{TV}(z)=P_{TV}(1-z),~~~P_{VT}(z)=P_{TT}(1-z),~~~~~~~z < 1.
\ee
In the leading order the kernel $P_{TV}(z)$ has a meaning of variation per unit
transfer momentum of the probability
density of finding a tensor-boson inside the vector-boson, $P_{VT}(z)$ - of finding
a vector inside
the tensor and $P_{TT}(z)$ - of finding a tensor inside the tensor.
For completeness we shall present also quark and vector-boson kernels
\cite{Dokshitzer:1977sg,Lipatov:1974qm,Altarelli:1977zs,Gribov:1972ri,Gribov:1972rt}:
\beqa\label{setofquarkgluon}
P_{qq}(z) &=& C_2(R){1+z^2 \over 1-z },\nn\\
P_{Vq}(z) &=& C_2(R){1+(1-z)^2 \over z },\\
P_{qV}(z) &=& T(R)[z^2 +(1-z)^2], \nn\\
P_{VV}(z) &=&  C_2(G)\left[{1 \over z(1-z)}+ {z^4 \over z(1-z)}+{(1-z)^4 \over z(1-z)}\right],\nn
\eeqa
where $C_2(G)= N, C_2(R)={N^2-1  \over  2 N},  T(R) = {1  \over  2}$ for the SU(N) groups.

{\it Having in hand the new set of splitting probabilities
for tensor-bosons  (\ref{setoftensorgluon})
we can consider a possible generalisation of quantum chromodynamics \cite{Savvidy:2013zwa}.
In so generalised theory in addition to the quarks and gluons there should be
 tensorgluons.  We can hypothesise that a possible emission of tensorgluons
by gluons, as it is shown on Fig.\ref{fig1},\ref{fig3} should produce a non-zero 
density of tensorgluons
inside the proton  in additional to the quark and gluon densities.
Our next goal is to derive DGLAP equations
\cite{Altarelli:1977zs,Gribov:1972ri,Gribov:1972rt,Lipatov:1974qm,Fadin:1975cb,Kuraev:1977fs,
Balitsky:1978ic,Dokshitzer:1977sg} which will take into account these new emission processes.}

\section{\it  Generalization of DGLAP Equation.\\
Calculation of Callan-Simanzik Beta Function }

In this section we shall  consider a possibility that inside the proton
and, more generally, inside hadrons there are
additional partons - tensorgluons, which can carry a part of the
proton momentum \cite{Savvidy:2014hha,Savvidy:2013zwa,Savvidy:2014uva}.
Tensorgluons have
zero electric charge, like gluons, but have a larger spin. Inside the proton
a nonzero density of the tensorgluons can be generated by the emission of tensorgluons
by gluons \cite{Savvidy:2005fi,Savvidy:2005zm,Savvidy:2005ki,Savvidy:2010vb}.
The last mechanism is typical for non-Abelian tensor gauge
theories, in which there exists a gluon-tensor-tensor vertex of order g
(see Fig.\ref{fig1}-\ref{fig2}) \cite{Savvidy:2005fi,Savvidy:2005zm,Savvidy:2005ki,Savvidy:2010vb}.
Therefore a number of gluons changes not only
because a quark may radiate a gluon or because a gluon may split into a quark-antiquark
pair or into two gluons \cite{Gross:1973ju,Gross:1974cs,Altarelli:1977zs},
but also because a gluon can split into two tensorgluons
\cite{Savvidy:2005fi,Savvidy:2005zm,Savvidy:2005ki,Savvidy:2010vb,Georgiou:2010mf,Antoniadis:2011rr}.
The process of gluon splitting into tensorgluons suggests that part of the proton momentum
which was carried by neutral partons can be  shared between vector and tensorgluons.
Our aim is to calculate the scattering amplitudes and splitting function in QCD generalised in this way.

It is well known that the deep inelastic structure functions
can be  expressed in terms of quark distribution densities.
If $q^i(x)$ is the density of quarks of type i (summed over
colours) inside a proton target with fraction x of the proton longitudinal momentum
in the infinite momentum frame \cite{Bjorken:1969ja} then the scaling structure functions can
be represented in the following form:
\be
2F_1(x)= F_2(x)/x= \sum_i Q^2_i [q^i(x)+\bar{q}^i(x)].
\ee
The scaling behaviour of the structure functions is broken and the results
can be formulated by assigning  a well determined $Q^2$
dependence to the parton densities.  This can be achieved by
introducing integro-differential equations which describe the
$Q^2$ dependence of quark  $q^i(x,t)$ and gluon densities  $G(x,t)$,
where $t=\ln(Q^2/Q^2_0)$ \cite{Altarelli:1977zs,Gribov:1972ri,Gribov:1972rt,
Lipatov:1974qm,Fadin:1975cb,Kuraev:1977fs,
Balitsky:1978ic,Dokshitzer:1977sg}.

Let us see what will happen if one supposes that
there are additional partons - tensorsgluons - inside the proton.
In accordance with our hypothesis there
is  an additional emission of tensorgluons in the proton, therefore one should introduce
the corresponding density $T(x, t)$ of tensorgluons (summed over colours)
inside the proton in the $P_{\infty}$ frame \cite{Bjorken:1969ja}. We can  derive
integro-differential equations that describe the $Q^2$ dependence
of parton densities in this general case \cite{Savvidy:2013zwa}:
\beqa\label{evolutionequation}
{d q^i(x,t)\over dt} &=& {\alpha(t) \over 2 \pi} \int^{1}_{x} {dy \over y}[\sum^{2 n_f}_{j=1} q^j(y,t)~
P_{q^i q^j}({x \over y})+ G(y,t)~ P_{q^i G}({x \over y})] ,\\
{d G(x,t)\over dt} &=& {\alpha(t) \over 2 \pi} \int^{1}_{x} {dy \over y}[\sum^{2 n_f}_{j=1} q^j(y,t)~
P_{G q^j}({x \over y})+ G(y,t) ~P_{G G}({x \over y})+ T(y,t) ~P_{G T}({x \over y}) ],\nn\\
{d T(x,t)\over dt} &=& {\alpha(t) \over 2 \pi} \int^{1}_{x} {dy \over y}[
G(y,t)~ P_{T G}({x \over y}) +  T(y,t)~ P_{T T}({x \over y})].\nn
\eeqa
The $\alpha(t)$ is the running coupling constant ($\alpha = g^2/4\pi$).
In the leading logarithmic approximation $\alpha(t)$ is of the form
\be\label{strongcouplingcons}
{\alpha \over \alpha(t)} = 1 +b ~\alpha ~t~~,
\ee
where $\alpha = \alpha(0)$ and $b$ is the one-loop Callan-Simanzik coefficient,
which, as we shall see below, receives an additional contribution from the tensorgluon
loop.
Here the indices i and j run over quarks and antiquarks of all flavors. The number
of quarks of a given fraction of momentum changes when a quark looses momentum by
radiating a  gluon,
or a  gluon inside the proton may produce a quark-antiquark pair \cite{Altarelli:1977zs}.
Similarly the number of  gluons changes
because a quark may radiate a gluon or because a gluon may split into a quark-antiquark
pair or into two gluons or {\it into two tensorgluons}. This last possibility is realised,
because, as we have seen, in non-Abelian tensor gauge
theories there is a triple vertex VTT (\ref{1ssvertex}) of a gluon and two tensorgluons of order g
\cite{Savvidy:2005fi,Savvidy:2005zm,Savvidy:2005ki,Savvidy:2010vb}.
This interaction should be taken into consideration,  and we added the
term $T(y,t) ~P_{G T}({x \over y})$ in the second
equation (\ref{evolutionequation}). The density of tensorgluons $T(x,t)$  changes when
a gluon splits into two tensorgluons or when a tensorgluon radiates a gluon. This
evolution is described by the last equation (\ref{evolutionequation}).

 In order to guarantee that the total momentum of the proton, that is, of
all partons is unchanged, one should impose the following constraint:
\be\label{conservation}
{d\over dt}\int_{0}^{1} dz z [\sum^{2n_f}_{i=1}q^{i}(z,t)+G(z,t)+T(z,t)]=0.
\ee
Using the evolution equations (\ref{evolutionequation}) one can express the derivatives
of the densities in (\ref{conservation}) in terms of kernels and to see that the following
momentum sum rules should be fulfilled:
\beqa\label{momentumsum}
&&\int_{0}^{1} dz z [P_{qq}(z)+P_{Gq}(z) ]=0,\nn\\
&&\int_{0}^{1} dz z [2 n_f P_{qG}(z)+P_{GG}(z)+P_{TG}(z)]=0,\nn\\
&&\int_{0}^{1} dz z [ P_{GT}(z)+P_{TT}(z)]=0.
\eeqa
Before analysing these momentum sum rules let us first
inspect the behaviour of the gluon-tensorgluon kernels (\ref{setoftensorgluon})
at the end points $z=0,1$. As one can see,
they are singular at the boundary values similarly to the
case of the standard kernels (\ref{setofquarkgluon}).
Though there is a difference here: the singularities are of higher order compared to the standard case
\cite{Altarelli:1977zs}.
Therefore one should define the regularisation procedure for the singular factors
$(1 - z)^{-2s+1}$ and $ z^{-2s+1}$  reinterpreting them as the  distributions $(1 - z)^{-2s+1}_{+}$ and
$z^{-2s+1}_{+}$, similarly to the Altarelli-Parisi regularisation \cite{Altarelli:1977zs}.
We shall define them  in the following
way:
\beqa\label{definition}
\int_{0}^{1} dz {f(z)\over (1 - z)^{2s-1}_+}&=&
\int_{0}^{1} dz {f(z)- \sum^{2s-2}_{k=0} {(-1)^k \over k!} f^{(k)}(1) (1-z)^k \over (1 - z)^{2s-1}},\nn\\
\nn\\
\int_{0}^{1} dz {f(z)\over z ^{2s-1}_+}&=&
\int_{0}^{1} dz {f(z)- \sum^{2s-2}_{k=0} {1 \over k!} f^{(k)}(0) z^k \over z^{2s-1}},\\
\nn\\
\int_{0}^{1} dz {f(z)\over z_+ (1-z)_+}&=&
\int_{0}^{1} dz {f(z)-  (1-z)f(0) - z f(1) \over z  (1-z) },\nn
\eeqa
where $f(z)$ is any test function which is sufficiently regular at the end points
and, as one can see, the defined substraction guarantees the convergence of the integrals.
Using the same arguments as in the standard case \cite{Altarelli:1977zs} we should add the delta function
terms into the definition of the diagonal kernels so that they will completely determine
the behaviour of $P_{qq}(z)$ , $P_{GG}(z)$ and $P_{TT}(z)$ functions. The first equation
in the momentum sum rule (\ref{momentumsum})
remains unchanged because there is no tensorgluon contribution into the quark evolution. The second
equation in the momentum sum rule (\ref{momentumsum}) will take the following form:
\beqa\label{betacoefficient}
&\int_{0}^{1} dz z [2 n_f P_{qG}(z)+P_{GG}(z)+P_{TG}(z) + b_G \delta (z-1)]=\nn\\
&=\int_{0}^{1} dz z  [2 n_f T(R)[z^2 +(1-z)^2]+C_2(G)\left[{1 \over z(1-z)}
+ {z^4 \over z(1-z)}+{(1-z)^4 \over z(1-z)}\right]+\nn\\
&+C_2(G)\left[ {z^4 \over z(1-z)}\left( {z\over 1-z} \right)^{2s-2}
+{(1-z)^4 \over z(1-z)} \left( {1-z\over  z}\right)^{2s-2} \right]  ] +b_G=\nn\\
&={2\over 3} n_f T(R) - {11 \over 6}C_2(G)- {12 s^2 -1\over 6} C_2(G) + b =0.
\eeqa
From this result we can extract an additional contribution
to the one-loop Callan-Symanzik beta function
arising from the tensorgluon loop.
Indeed, the first beta-function coefficient enters into this expression because the momentum sum
rule (\ref{momentumsum}) implicitly comprises unitarity, thus the one-loop effects \cite{Altarelli:1977zs}.
In (\ref{betacoefficient}) we have three terms which come from gluon and quark loops:
\be
b_{1} = {11 \over 6}C_2(G) -{  2 n_f   \over 3}T(R),
\ee
and  from the tensorboson loop of spin s:
\be\label{spinscontr}
b_T =  {12 s^2 -1\over 6} C_2(G), ~~~s=1,2,3,4,....
\ee
It is a very interesting result because at s=1 we are rediscovering the
asymptotic freedom result \cite{Gross:1973ju,Gross:1974cs,Politzer:1973fx}.
For larger spins the  tensorgluon contribution into the Callan-Simanzik beta function
has the same signature as the standard gluons, which means that tensorgluons
"accelerate" the asymptotic freedom (\ref{strongcouplingcons}) of the strong
interaction coupling constant $\alpha(t)$.
The contribution is increasing quadratically
with the spin of the tensorgluons, that is, at large transfer momentum
the strong coupling constant tends to zero faster compared to the standard case:
\be
\alpha(t)= {\alpha \over 1+ b  \alpha ~t }~,
\ee
where
\be\label{fullbeta}
b =  {(12s^2 -1) C_2(G) - 4 n_f T(R) \over 12 \pi},~~~~~s=1,2,...
\ee
Surprisingly, a similar result based on the parametrization of the
charge renormalization taken in the form  $b = (-1)^{2s}(A+Bs^2)$
was conjectured  by Curtright
\cite{Curtright:1981wv}. Here $A$ represents an orbital contribution and
$B s^2$ - the anomalous magnetic moment contribution
\cite{Savvidy:1977as,Matinyan:1976mp,Batalin:1976uv}. The unknown coefficients A and B were found
by comparing the suggested parametrisation with the known results for s= 0, 1/2 and 1.

It is also possible to consider a straitforward generalisation of the result obtained
for the effective action in Yang-Mills theory long ago
\cite{Savvidy:1977as,Matinyan:1976mp,Batalin:1976uv,Kay:1983mh}
to the higher spin gauge bosons. With the spectrum of the tensorgluons in
the external chromomagnetic field  $\lambda = (2n+1 + 2s)gH +k^2_{\parallel}$
one can perform a summation of the modes and get an exact result for the one-loop effective action similarly to \cite{Savvidy:1977as,Kay:1983mh}:
\be
\epsilon= {H^2 \over 2} +{(gH)^2 \over 4\pi} ~b ~[\ln{gH \over \mu^2}-{1\over 2}],
\ee
where
\beqa\label{savv}
b =  -{2 C_2(G)\over \pi} ~ \zeta(-1, {2s+1\over 2})=
{12s^2-1\over 12 \pi}C_2(G),
\eeqa
and $\zeta(-1, q)=-{1\over 2}(q^2 -q +{1\over 6})$ is the generalised zeta
function\footnote{The generalised zeta function is defined as $\zeta(p, q)=\sum^{\infty}_{k=0}{1\over (k+q)^p}
={1\over \Gamma(p)} \int^{\infty}_{0} dt t^{-1+p} { e^{-qt} \over 1-e^{-t}} $ .}.
Because the coefficient in front of the logarithm defines  the beta function
\cite{Savvidy:1977as,Matinyan:1976mp}, one can see that (\ref{savv}) is
in agrement with the result (\ref{spinscontr}).

It is also natural to ask what will happen if one takes into consideration the contribution
of tensorgluons of all spins into the beta function\footnote{I would like to thank
John Iliopoulos and Constantin Bachas for raising this question.}. One can suggest two scenarios.
In the first one the high spin gluons, let us say, of $s  \geq 3$, will get large mass
and therefore they can be ignored at a given energy scale. In the second case,
when all of them remain massless, then one can
suggest the Riemann zeta function regularisation, similar to the Brink-Nielsen regularisation
\cite{Brink:1973kn}. The summation over the spectrum in (\ref{spectrum1}) gives\cite{Savvidy:2014uva}:
\beqa
b_{11} &=& C_2(G) [ \sum^{\infty}_{s=1} {( 12s^2 -1) \over 12 \pi}+
\sum^{\infty}_{s=0} {( 12s^2 -1) \over 12 \pi} +
\sum^{\infty}_{s=1} {( 12s^2 -1) \over 12 \pi}+
\sum^{\infty}_{s=0} {( 12s^2 -1) \over 12 \pi} +.....]\nn\\
&=&C_2(G)[{1\over \pi}  \zeta(-2)- {1\over 12\pi} \zeta(0)-{1\over 12\pi}
+{1\over \pi}  \zeta(-2)- {1\over 12\pi} \zeta(0) +...] \nn\\
&=&C_2(G)[{1\over 24\pi} - {1\over 12\pi} +
{1\over 24\pi}+...]=0,
\eeqa
where $\zeta(-2)=0,~ \zeta(0)=-1/2$, leading to the theory which is  {\it conformally invariant} at very high energies. The above
summation requires explicit regularisation and further justification.

\section{\it Unification of Coupling Constants of Standard Model}

It is interesting to know how the contribution of tensorgluons changes the high energy
behaviour of the coupling constants of the Standard Model \cite{Georgi:1974sy,Georgi:1974yf}.
The coupling constants are evolving in accordance with the formulae
\be\label{system}
{1 \over \alpha_i(M)} = {1 \over \alpha_i(\mu)}+  2 b_i  \ln{M\over \mu},~~~i=1,2,3,
\ee
where we shall consider only the contribution of the lower $s=2$ tensorbosons:
\be
2b = {58 C_2(G) - 4 n_f T(R) \over 6 \pi}.
\ee
For the $SU(3)_c \times SU(2)_L \times U(1)$  group with its coupling constants $\alpha_3, \alpha_2$ and $\alpha_1$
and six quarks $n_f=6$ and $SU(5)$ unification group we will get
$$
2 b_3= {1 \over 2\pi} 54,~~~ 2 b_2= {1\over 2 \pi} {104\over 3},~~~2 b_1= -{1\over 2 \pi} 4,
$$
so that solving the system of equations (\ref{system}) one can get
\be
\ln{M\over \mu} = {\pi \over 58} \left({1\over \alpha_{el}(\mu)}- {8\over 3} {1\over \alpha_s(\mu)}\right),
\ee
where $\alpha_{el}(\mu)$ and $\alpha_s(\mu)$ are the electromagnetic and strong coupling constants at
scale $\mu$. If one takes $\alpha_{el}(M_Z)= 1/128$ and $\alpha_s(M_Z) =1/10$ one can get that
coupling constants have equal strength at energies of order
$$
M \sim 4 \times 10^4 GeV = 40~ TeV,
$$
which is much smaller than the scale $M\sim  10^{14} GeV$  in the absence of the tensorgluons contribution.
The value of the weak angle \cite{Georgi:1974sy,Georgi:1974yf} remains intact :
\be
\sin^2\theta_W =  {1\over 6} +{5\over 9} {\alpha_{el}(M_Z)\over \alpha_s(M_Z)},
\ee
as well as the coupling constant at the unification scale remains of the same order
 $\bar{\alpha}(M)=0,01$.

\section{\it Conclusion}

In the present article we describe a possible extension  
the Yang-Mills gauge principle \cite{yang} which includes tensor gauge fields.
In this extension of the Yang-Mills theory the vector gauge boson becomes a
member of a bigger family of gauge bosons of arbitrary large integer spins.

The proposed extension of Yang-Mills theory 
is essentially based on the existence of the enlarged  Poincar\'e 
algebra and on an appropriate transversal representations
of that algebra.  
The invariant Lagrangian is expressed in terms of new
higher-rank field strength tensors. The Lagrangian  does not contain higher derivatives of
tensor gauge fields and all interactions take place through three- and four-particle
exchanges with a dimensionless coupling constant (see Fig.\ref{fig1}-\ref{fig2}).

We calculated the scattering amplitudes of non-Abelian tensor gauge bosons
at tree level, as well as their one-loop contribution into the Callan-Symanzik beta function.
This contribution is negative and
corresponds to the asymptotically free theory. The proposed extension may
lead to a natural inclusion of the standard theory of fundamental forces
into a larger theory in which vector gauge bosons, leptons and quarks represent
a low-spin subgroup. 

In the line with the above development we considered a possible extension of QCD. 
In so extended QCD inside the proton and, more generally, inside hadrons there should be
additional partons - tensorgluons, which can carry a part of the proton
momentum.  Among all parton distributions, the gluon density $G(x,t)$ is one of the least
constrained functions since it does not couple directly to the
photon in deep-inelastic scattering measurements of the proton $F_2$ structure function.
Therefore it is only indirectly constrained by scaling violations and by the momentum sum rule
which resulted in the fact that only half of the proton momentum is carried by charged
constituents - the quarks - and that the other part is ascribed to the
neutral constituents.

As it was suggested, the process of gluon splitting leads to the emission
of tensorgluons and
therefore a part of the proton momentum which is carried  by the neutral constituents
can be  shared between gluons and tensorgluons.  The density of neutral partons in the proton is
therefore given by the sum of two functions: $G(x,t)+T(x,t)$, where $T(x,t)$
is the density of the
tensorgluons. To disentangle these contributions and to decide which 
piece of the neutral partons is the contribution of gluons and which one
is of the tensorgluons one should measure the helicities of the neutral
components, which seems to be a difficult task.

The gluon density can be directly constrained by jet production \cite{Ellis:1976uc}.
In the suggested model the situation is such that the standard quarks 
cannot radiate tensorgluons
(such a vertex is absent in the model
\cite{Savvidy:2005fi,Savvidy:2005zm,Savvidy:2005ki,Savvidy:2010vb}),
therefore only gluons are radiated by quarks.
A radiated gluon then can split into a pair of tensorgluons 
without obscuring the structure of the
observed three-jet final states.  Thus it seems that there is no obvious contradiction with the existing
experimental data.  Our hypotheses may be wrong, but the uniqueness and simplicity of
suggested extension seems to be the reasons for serious consideration.

This extension of QCD influences the unification scale at which
the coupling constants of the Standard Model merge.
In the last section we observed that the unification scale at which
standard coupling constants are
merging is  shifted to lower energies telling us that it may be that a
new physics is round the corner.
Whether all these phenomena are consistent with experiment is an open question.

\section{\it Acknowledgement }
I would like to thank the organisers of the "Conference on 60 Years of Yang-Mills Gauge Field Theories"  for their kind hospitality in the Institute of Advanced Studies of the 
Nanyang Technological University 
in  Singapore and Prof. Kok-Khoo Phua and Prof. Yong Min Cho for invitation.
This work was supported in part by the General Secretariat for Research and
Technology of
Greece and the European Regional Development Fund (NSRF 2007-15 ACTION,KRIPIS).

\end{document}